\documentclass[iop]{emulateapj}
\usepackage{apjfonts}
\usepackage{graphicx}
\usepackage{natbib}
\usepackage{amsmath}
\usepackage{color,soul}
\usepackage{comment}
\usepackage[version=3]{mhchem}
\begin{document}
\title{Scaling the Earth: A Sensitivity Analysis of Terrestrial Exoplanetary Interior Models}
\author{C. T. Unterborn\altaffilmark{*}, E. E. Dismukes and W. R. Panero}
\affil{School of Earth Sciences, The Ohio State University, 125 South Oval Mall, Columbus, OH 43210}
\altaffiltext{*}{contact: unterborn.1@osu.edu}
\received{October 26, 2015}
\accepted{January, 7, 2016}

\begin{abstract}
\indent An exoplanet's structure and composition are first-order controls of the planet's habitability.  We explore which aspects of bulk terrestrial planet composition and interior structure affect the chief observables of an exoplanet: its mass and radius. We apply these perturbations to the Earth, the planet we know best. Using the mineral physics toolkit BurnMan to self-consistently calculate mass-radius models, we find that core radius, presence of light elements in the core and an upper-mantle consisting of low-pressure silicates have the largest effect on the final calculated mass at a given radius, none of which are included in current mass-radius models. We expand these results provide a self-consistent grid of compositionally as well as structurally constrained terrestrial mass-radius models for quantifying the likelihood of exoplanets being ``Earth-like.'' We further apply this grid to Kepler-36b, finding that it is only $\sim$20\% likely to be structurally similar to the Earth with Si/Fe = 0.9 compared to Earth's Si/Fe = 1 and Sun's Si/Fe = 1.19. \end{abstract}

\section{Introduction}
\label{sec:intro}
\indent  For characterizing extrasolar planets, the current observational techniques offer only two parameters for constraining a planet's interior composition: its mass and radius. Of the $\sim$1600 confirmed planets discovered to date, 69\% have both mass and radius measurements available from which mean density is calculated \citep{ExoOrg}. The mean density of a planet, in turn, is a function of its composition \citep{Zapo69}. Models of varying proportion of metal, rock, water, and H$_{2}$/He suggest the existence of planets potentially being terrestrial ``super Earths,'' gaseous ``mini Neptunes,'' water (vapor) planets, or gas giants dominated by H$_{2}$ and He. This compositional and mass diversity then provides us with a multifaceted dataset with which to examine the extent of interior planet compositions within our Galaxy. \\
\indent The smallest planets are terrestrial, in which a transparent atmosphere accounts for a negligible fraction of the planetary mass. Studies examining the specific terrestrial systems Kepler-10b \citep{Bata11}, CoRoT-7b \citep{Quel09,Hatz11}, Kepler-36b \citep{Cart12}, as well as the more general case of interior composition's effect on observed mass and radius \citep{Vale06,Vale07a,Vale07b,Fort07,Sot07,Gras09,Seag07,Wagn11,Swif12,Weis14,Lope14} all point to density degeneracy in interior composition and structure from only these observables, independent of the large uncertainties present in the observational measurement \citep{Weis14}.\\
\indent In part because of this degeneracy, most existing mass-radius models for terrestrial exoplanets \citep[e.g.][]{Seag07, Zeng13} simplify the complex mineralogy and structure within terrestrial planets by adopting a layered model of a pure, solid Fe-core with a magnesium-silicate mantle, as these two factors dominate the average density of the planets. \citet{Dorn15} expand this simple layered model by adopting stellar Fe, Mg and Si abundances as proxies for bulk planetary composition in order to further constrain aspects of interior structure, particularly the relative size of a planet's core and mantle. This assumption is well grounded for our Solar System, with both Earth and chondritic abundances being nearly identical to that of the Solar photosphere for these refractory, planet-building elements \citep{McD03,Lodd03}. \\
\indent As we seek to characterize ``Earth-like'' planets, then, it is reasonable to evaluate the degree to which these current models reproduce the Earth's structure in an appropriate manner so as to draw appropriate conclusions on its dynamic state and potential habitability. The Earth, has a complex internal structure, in which light elements, likely dominated by Si, make up between 5-10\% of the core by weight \citep{Fisch15}, relatively dense Fe-oxides and Fe-silicates make up 10-12\% of the mantle \citep{Irif98}, with a series of pressure-induced phase transitions in the mantle \citep{Jean83}, and a melt-extracted, low density crust \citep{Rudn03}. All of these factors result from an early chemical differentiation processes likely occurring as well in any terrestrial exoplanet whose composition is dominated by iron and oxides, none of which are addressed currently in mass-radius models. \\
\indent Simplified two-layer planet models with a core mass fraction identical to the Earth, over predict the Earth's total mass by $\sim$13\% \citep{Zeng13}. Applying such a simplified model to the Earth would therefore predict a smaller core. Such a difference will erroneously lead to a planet model with more rapid heat loss from the core, limiting the lifetime of a magnetic field and affecting the viability of plate tectonics at the surface assuming a heat budget equal to that of the Earth. These two dynamic processes are likely required to sustain liquid water and life on Earth \citep{Fole15,Dris15}. While still within the uncertainty of most planet mass measurements, as observational techniques improve, the consequences of these ``spherical cow'' models will become more and more evident.  \\
\indent Here, we present a systematic sensitivity analysis of how uncertainties in material properties, core composition, major element chemistry (Mg, Fe, Si, O), and structural transitions in mantle silicates affect the mean density of an Earth-sized, terrestrial planet. In section 2 we describe the mass-radius model employing a self-consistent thermoelastic compression calculator, BurnMan \citep{Cott14}\footnote{available at www.burnman.org}. In section 3 we present the resulting density and variance in density resulting from changing these individual parameters. Finally in sections 4 and 5, we discuss the level of chemical and structural complexity necessary to consider within a planetary interior model to adequately infer the Earth's composition despite the degeneracy in the solution space of the forward model. We then apply these results to exoplanet Kepler-36b and provide predictions for the stellar composition its host star. 

\begin{deluxetable}{lccccc}
\tablecolumns{6}
\tabletypesize{\scriptsize}
\tablewidth{0pt}
\tablecaption{Thermoelastic parameters adopted in our mass-radius model}
\tablehead{\colhead{}&\colhead{V$_{0}$}&\colhead{$\rho_{0}$}&\colhead{K$_{0}$}&\colhead{ }&\colhead{ }\nl\colhead{Mineral}&\colhead{(cm$^{3}$ mol$^{-1}$)}&\colhead{(g cm$^{-3}$)}&\colhead{(GPa)}&\colhead{K$^{'}$}&\colhead{EOS}}
\startdata
Liquid-Fe$^{\sharp}$&7.96&7.019&109.7&4.66&BM4\nl
$\epsilon$-Fe$^{\mathsection}$&6.75&8.27&163.4&5.38&Vinet\nl
$\epsilon$-Fe$^{\dagger}$&6.72&8.31&156.2&6.08&Vinet\nl
MgSiO$_{3}$$^{\ddagger}$&24.45&4.11&251&4.1&SLB3$^{*}$\nl
\multicolumn{6}{l}{(brigmanite)}\nl
MgO$^{\ddagger}$&11.24&3.59&161&3.8&SLB3$^{*}$\nl
Mg$_{2}$SiO$_{4}$$^{\ddagger}$&43.6&3.23&128&4.2&SLB3$^{*}$\nl
MgSiO$_{3}$$^{\ddagger}$&31.4&3.2&107&7.0&SLB3$^{*}$\nl
\multicolumn{6}{l}{(enstatite)}\nl
FeSiO$_{3}$$^{\ddagger}$&25.49&5.18&272&4.1&SLB3$^{*}$\nl
\multicolumn{6}{l}{(pv structure)}\nl
FeO$^{\ddagger}$&12.26&5.86&179&4.9&SLB3$^{*}$\nl
\enddata
\tablerefs{\scriptsize  $^{\mathsection}$\citet{Dewa00} $^{\dagger}$\citet{Ande01} $^{\sharp}$ \citet{Ander94}; 4th Order Birch-Murnaghan EOS with K$^{\prime \prime}$ = -0.043*10$^{-9}$ GPa$^{-1}$ along an isentrope centered at 1 bar and 1811 K $^{\ddagger}$\citet{Stix05} and references therein $^{*}$While the \citet{Stix05} EOS formulation is a thermal EOS, it reduces to the third-order Birch-Murnaghan EOS at 300 K. We therefore adopt a geotherm of constant $T$ = 300 K.}
\label{parameters} 
\end{deluxetable}
\section{Methods}
Planetary mass-radius models are calculated by solving three coupled, differential equations for the mass within a sphere,
\begin{equation}
\label{MR1}
\frac{dm(r)}{dr}=4\pi r^{2}\rho(r)
\end{equation}
\noindent{the equation of hydrostatic equilibrium,}
\begin{equation}
\label{MR2}
\frac{dP(r)}{dr}=\frac{-Gm(r)\rho(r)}{r^2}
\end{equation}
\noindent{and the equation of state,}
\begin{equation}
\label{MR3}
P(r)=f(\rho(r),T(r))
\end{equation}
where $r$ is the radius, $m(r)$ is the mass within a shell of radius $r+dr$, $\rho$ is the density, $P$ is the pressure, $T$ is the temperature, and $G$ is the gravitational constant. All calculations integrated from the planetary surface, at which $P_{r=0}=0$, to the center. With central pressure treated as a free parameter, we are able to focus on the structural variability constrained by planetary radius measured from transits, in which the final mass is a model result of our choice of input parameters. This is in contrast to other mass-radius models which assume a central pressure and integrate upward to a pressure cutoff of $P$=0.\\
\indent In the high temperature limit, the thermal expansion coefficient at constant pressure, $\alpha$, for mantle silicates is essentially constant. This coefficient is defined as:
 \begin{equation}
\label{thermal}
\alpha = \frac{1}{V}\left (\frac{\partial V}{\partial T}\right )_{P}
\end{equation}
where $V$ is the planetary volume. Adopting a value $\alpha$ characteristic of silicate minerals ($\alpha \sim 10^{-6}$), we see that even for differences in temperature on the order $\partial T \sim 10^{4}$, the change in the planet's volume is only on the percent level ($\partial V \approx 10^{-2}\ast V$). We therefore adopt for these mass-radius calculations an isothermal EOS for all calculations ($T$ = 300 K).\\
\indent Equations \ref{MR1}-\ref{MR3} are solved using the open-source BurnMan code \citep{Cott14} adapted to include a Vinet and 4th-order Birch-Murnaghan EOS using thermoelastic data in Table \ref{parameters}. Few EOS parameters are measured at $P$-$T$ conditions beyond the Earth's core-mantle boundary \citep{Voca07}. Geochemical models predict variations in the bulk mineralogy of the mantle and the abundance and identity of light elements in the core \citep{McD03,Javo10}. Seismological models, backed by lab-based experiments, confirm multiple pressure-induced phase transitions in the silicate minerals of the mantle, most notably the transitions in the low pressure mantle silicates to periclase (MgO) + bridgmanite (MgSiO$_{3}$), with a net density increase of 14.8\%. With these factors in mind, we address the consequences of uncertainties in both the equation of state of the constituent planetary minerals as well as those regarding the mineralogy and bulk composition as a function of depth within mass-radius calculations. \\
\indent A further structural planetary constraint is the size of the core. For the Earth, the core mass and radius are not uncertain because of the availability of both seismic and moment of inertia data. However, no remote observation of an exoplanet can infer the core size directly, and we therefore address the consequence of core size as a free parameter. 
\begin{figure}
\begin{centering}
\includegraphics[width=6cm]{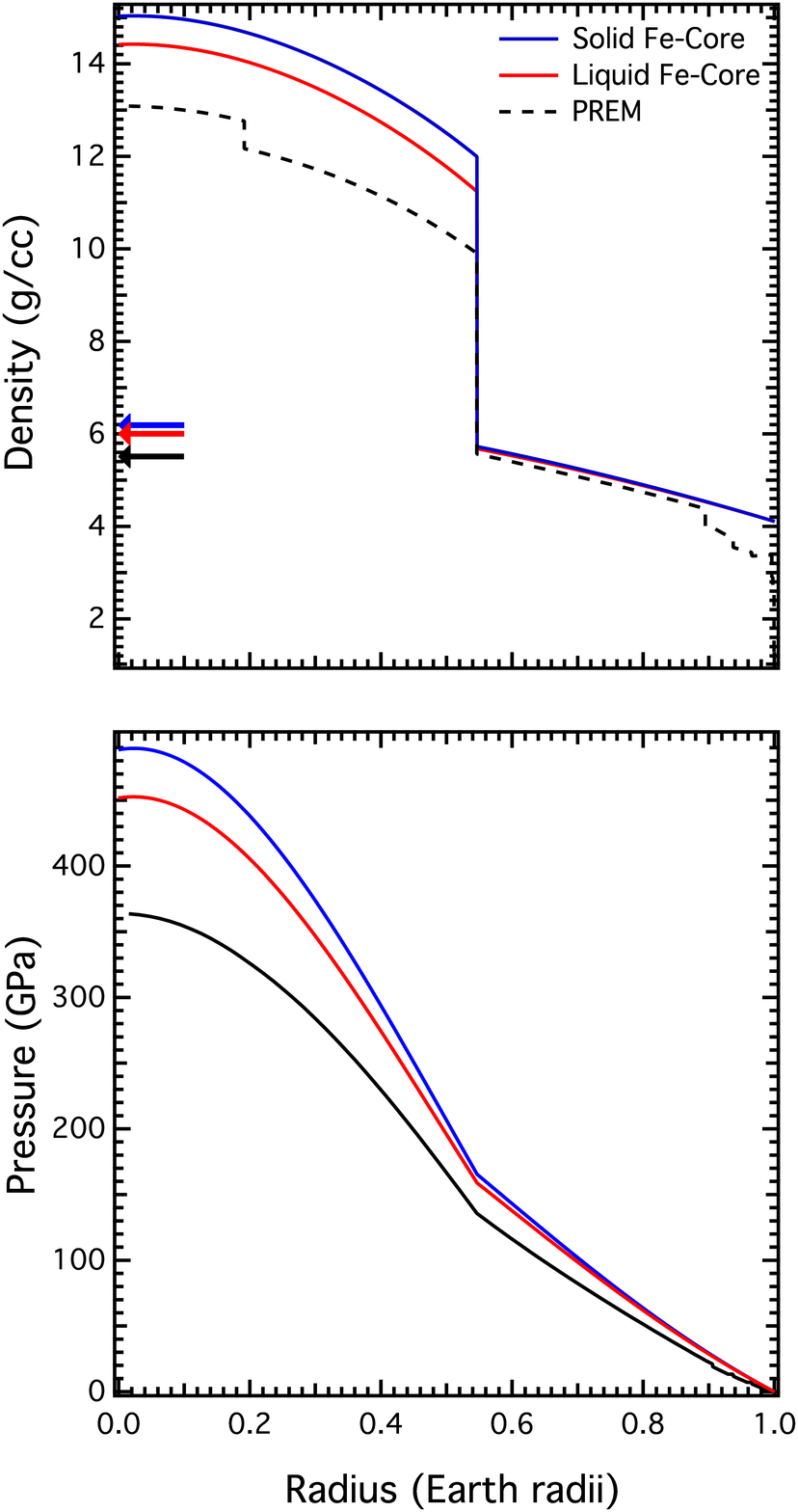}
\caption{Density and pressure profiles for a two-layer (pure Fe-core, MgSiO$_{3}$-bridgmanite mantle), one Earth radius planet as a function of planet radius for models adopting an entirely solid (blue) and liquid (red) Fe core. For comparison between \citet{Seag07} and \citet{Zeng13}, we adopt the solid-Fe EOS of \citet{Ande01} for this calculation. The Preliminary Reference Earth Model \citep[PREM, ][]{PREM} is shown in black-dashed for comparison. Average planet densities for each model is shown as arrows of same color scheme as density/pressure profiles.}
\label{Earth_mod}
\end{centering}
\end{figure}
\begin{figure}
\begin{centering}
\includegraphics[width=6cm]{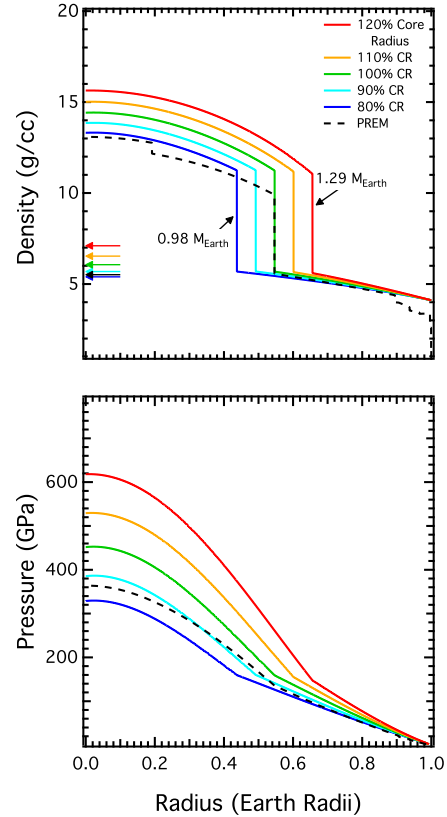}
\caption{Density and pressure profiles for a two-layer (pure, liquid Fe-core, MgSiO$_{3}$-bridgmanite mantle), one R$_{\oplus}$ planet as a function of planetary radius for variable core radius. PREM is shown in black-dashed for comparison. Average planet densities for each model is shown as arrows of same color scheme as density/pressure profiles. Figures \ref{CoreRad}-\ref{FeMantle} are all plotted on the same scale for comparison.}
\label{CoreRad}
\end{centering}
\end{figure}
\section{Results}
First, we calculate the density structure of a planet with an entirely solid Fe-core and perovskite-structured magnesium silicate, ``bridgmanite,'' mantle with a core radius equal to that of the Earth \citep[1 CRE = 3480 km, ][]{PREM} similar to the models of \citet{Seag07} and \citet{Zeng13}. This model over-predicts the Earth's mass by $\sim$13\% (Figure \ref{Earth_mod}) in agreement with these models for similar input parameters. Furthermore, this model predicts an Earth with Si/Fe = 1.01, almost precisely the Earth value of 1.0 \citep{McD03}. Within this two-layer, solid-Fe core model then, the overestimate in mass leads to the conclusion of the Earth's core being $\sim$20\% smaller with Si/Fe = 2.25, greater than 200\% of the Earth value. This choice of Fe EOS  \citep[][Table \ref{parameters}]{Dewa00} is derived from quasi-hydrostatic, room temperature data measured to 200 GPa using a Vinet EOS. We prefer this EOS for solid-Fe rather than the EOS data of \citet{Ande01} used in both \citet{Seag07} and \citet{Zeng13}, as the  \citet{Ande01} data is based on compression data to 330 GPa without a pressure medium, which leads to significant non-hydrostatic stresses and underestimate of density at a given pressure. In either case, however, the compression behavior of iron remains uncertain in the strictest sense, and must be extrapolated to the $P$-$T$ regime of Super-Earth's with caution.\\
\indent By both volume and mass, however, liquid-Fe is the dominant phase in the Earth's core. We therefore adopt as our reference model a planet with an entirely liquid Fe-core and bridgmanite mantle with a core radius equal to 1 CRE. While the density deficit between liquid and solid-Fe at these pressures is $\sim$0.4 g/cm$^{3}$, or 3\% \citep{PREM}, this drop does not account for the excess mass in the two-layer model (Figure \ref{Earth_mod}, red line). Furthermore, even accounting for the presence of liquid-Fe, one would still underestimate the core's radius by $\sim$20\% (Figure \ref{CoreRad}).
\subsection{Iron Equations of State}
The Earth's core makes up nearly 32\% of its total mass \citep{McD03} and spans a pressure range between 136 and 360 GPa \citep{PREM}. Our choice for Fe EOS (Table \ref{parameters}) is calculated along a isentrope centered at 1 bar and 1811 K \citep{Ander94}. In order to account for any possible error in our extrapolation to higher pressures then, we compare our reference model for a 1 Earth-radius planet (1 R$_{\oplus}$), with a 1 CRE sized liquid-Fe core, varying only the bulk modulus of Fe. We find that even with a 15\% increase in iron's bulk modulus, this model still over predicts the Earth's true mass by 9\% (Figure \ref{Modulus}a). Furthermore, we find only a $\sim$0.03 M$_{\oplus}$ difference between our high and low bulk modulus models, such that uncertainties in iron compressibility are a minor contributor to the overestimate of Earth mass in a simple two-layer model.
\subsection{Light elements in the core}
\begin{figure}
\begin{centering}
\includegraphics[width=8cm]{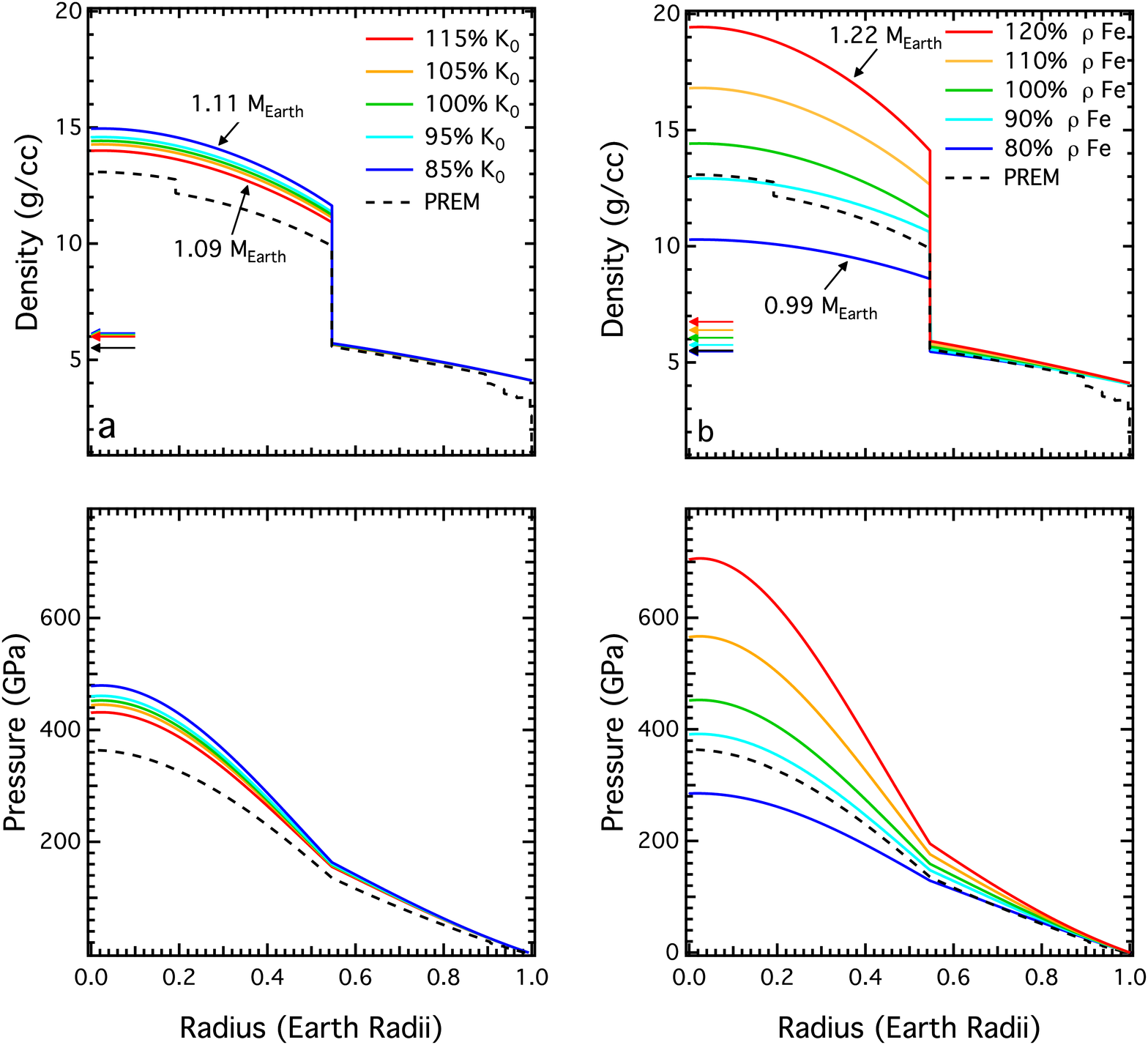}
\caption{Density and pressure profiles for a two-layer (pure Fe-core, MgSiO$_{3}$ mantle), one Earth radius planet of variable core bulk modulus (K$_{0}$ = 109.7 GPa) and density at 300 K, 1 bar ($\rho_{0}$ = 7.96 cm$^{3}$ mol$^{-1}$). PREM is shown in black-dashed for comparison. Average planet densities for each model is shown as arrows of same color scheme as density/pressure profiles. Figures \ref{CoreRad}-\ref{FeMantle} are all plotted on the same scale for comparison.}
\label{Modulus}
\end{centering}
\end{figure}
The Earth's core is less dense than that of pure Fe-alloy by between 5 and 10\%, suggesting that the core contains some lighter elements \citep{Birch52,Birch61,Jean79}. This incorporation of light elements is a consequence of low- to moderate-pressure partitioning of chemical species between iron and silicates during the differentiation of the planets during the formation process. To first order, this density reduction does not significantly affect compressibility of Fe, but reduces the molar mass of the core itself \citep{Poir94}. We therefore model the impact of light elements on planetary structure by scaling $\rho_{0}$ between 80 and 120\% (Figure \ref{Modulus}b) of pure Fe. This affects not only the total planet's mass, but also affects the pressure gradient within both the core and mantle. In the 120\% $\rho_{0}$ model, for example, the central pressure is more than double that observed in the Earth leading to a greater pressure at the base of the mantle, with addition mass excess contributed from the mantle due to compression. A planet of 1 M$_{\oplus}$ and R$_{\oplus}$ can be achieved with a 20\% core density reduction. This model underestimates the mass of the core (Figure \ref{Modulus}), but is compensated by the neglect of the phase transitions in mantle silicates, leading to serendipitous agreement.
\subsection{Mantle Mg/Si ratio}
\begin{figure}
\begin{centering}
\includegraphics[width=6cm]{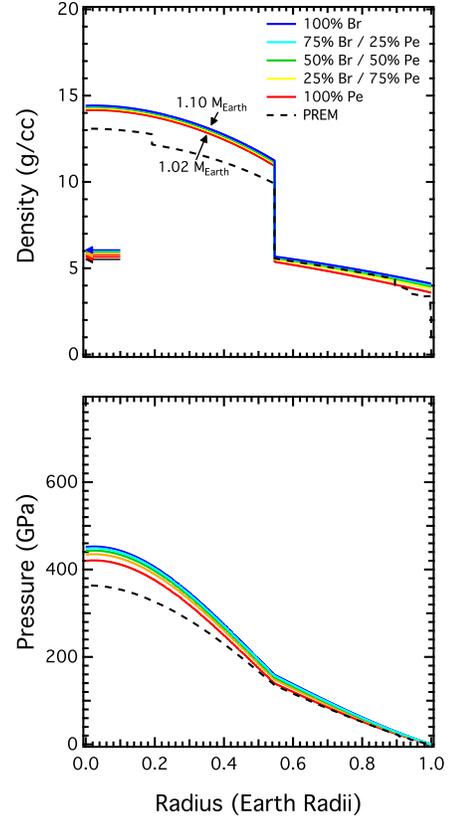}
\caption{Density and pressure profiles for a two-layer (pure Fe-core, MgSiO$_{3}$ mantle), 1 R$_{\oplus}$ planet of variable mantle composition between entirely brigmanite (br, MgSiO$_{3}$ and entirely periclase (pe, MgO). PREM is shown in black-dashed for comparison. Average densities for these models shown as arrows of same color scheme as density/pressure profiles. Figures \ref{CoreRad}-\ref{FeMantle} are all plotted on the same scale for comparison.}

\label{Mantle}
\end{centering}
\end{figure}

The silicate mineral brigmanite (perovskite-structured MgSiO$_{3}$) and periclase (MgO) account for $\sim$79\% of the Earth mantle mass. Varying proportions of each mineral has the effect of varying the planet's Mg/Si ratio as well as the Si/Fe ratio from 1.03 to 0, in the absence of Si in the core. As both refractory elements, this ratio is likely similar to the stellar ratio \citep{Lodd10}, and geochemical models predict between 80\% \citep{McD03} and $\sim$100\% brigmanite \citep{Javo10}, consistent with stellar Mg/Si$\sim$1.02 \citep{Lodd03}. We vary the proportion of each while constraining the core size to 1 CRE (Figure \ref{Mantle}) effectively varying the planetary Mg/Si ratio. Increasing the fraction of periclase within our model decreases total mass a small amount while decreasing Mg/Si. Even in the model with a 100\% MgO mantle, the simple two-layer model reduces planetary mass to 1.02 M$_{\oplus}$.
\subsection{FeO in the mantle}
\begin{figure}
\begin{centering}
\includegraphics[width=6cm]{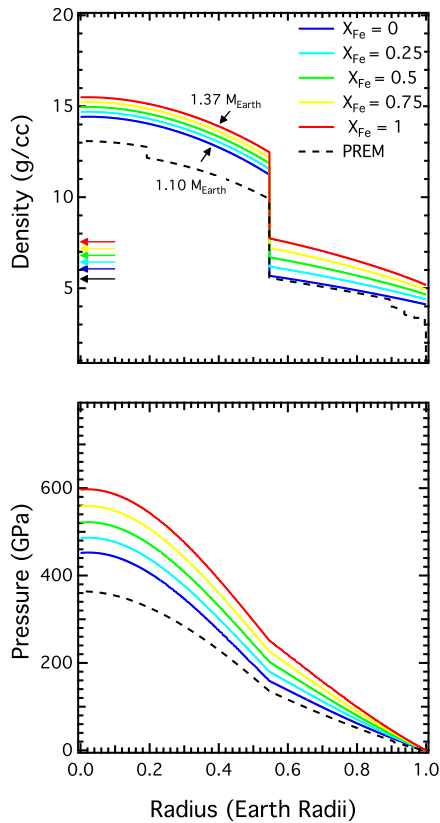}
\caption{Density and pressure profiles for a two-layer (pure Fe-core, (Mg$_{1-x}$,Fe$_{x}$)SiO$_{3}$ mantle), 1 R$_{\oplus}$ planet of varying Fe incorporation into the mantle between pure MgSiO$_{3}$ and pure FeSiO$_{3}$. PREM is shown in black-dashed for comparison. Average planet densities for each model is shown as arrows of same color scheme as density/pressure profiles. Figures \ref{CoreRad}-\ref{FeMantle} are all plotted on the same scale for comparison.}

\label{FeMantle}
\end{centering}
\end{figure}
The Earth contains 4.2\% by weight and $\sim$5.4\% by mol Fe in its mantle \citep{McD03}, a consequence of chemical partitioning of iron between silicate and metal in the planetary differentiation process. Fe is distributed between brigmanite and periclase as FeSiO$_{3}$ and FeO, respectively. Increasing the iron content of the mantle has negligible effect on compressibility of either mineral, however increases their densities due to the higher molecular weight compared to the magnesium end-members (Figure \ref{FeMantle}). We find that the small amount of Fe present in the Earth's mantle only accounts for a 2-3\% increase in mass relative to a pure-Mg mantle, and is thus a small factor in a planet's total mass compared to the other factors explored here. 
\subsection{Upper Mantle}
The Earth's mantle, however, is not entirely composed of the lower-mantle minerals considered here. Instead, at < 660 km depth or $P$ < 25 GPa, the lower mantle minerals are unstable, with olivine (Mg$_{2}$SiO$_{4}$) and pyroxenes (MgSiO$_{3}$) dominating. Upper-mantle phase transitions have a net density decrease of 14.8\%, while the compression due to pressure is only 8\%. Therefore phase transitions have nearly twice the impact on density as compression near the surface. The Earth's upper mantle makes up $\sim$17\% of its total mass, however, this mass fraction is proportionally less important for larger planets where the 25 GPa phase boundary is relatively shallower. For Earth-sized or smaller planets then, this density drop causes a non-trivial reduction in the total mass of the modeled planet and is currently not included in any of the available mass-radius models. Case in point, models of Mars interior structure place the core mantle boundary very near 25 GPa, such that nearly the entire mantle is composed of low-pressure minerals.  \\
\indent To model the presence of an upper mantle then, we adopt the EOS of a mixture of pyroxene and olivine for depths where the pressure is less than 25 GPa. For consistency, the exact proportion of olivine to pyroxene is determined by adopting the Mg/Si of the lower mantle. Thus, for a lower mantle containing 67\% brigmanite and 33\% periclase (Mg/Si = 1.5), the upper mantle would contain 50\% olivine and 50\% pyroxene. We find in the 1 CRE, 100\% brigmanite model, the total mass decreases from 1.10 M$_{\oplus}$ in the two-layer model to 1.04 M$_{\oplus}$ in the 3 layer model. This 6\% drop is applicable to all previous models when an upper mantle is included. 
\section{Discussion}
\begin{figure}
\begin{centering}
\includegraphics[width=8cm]{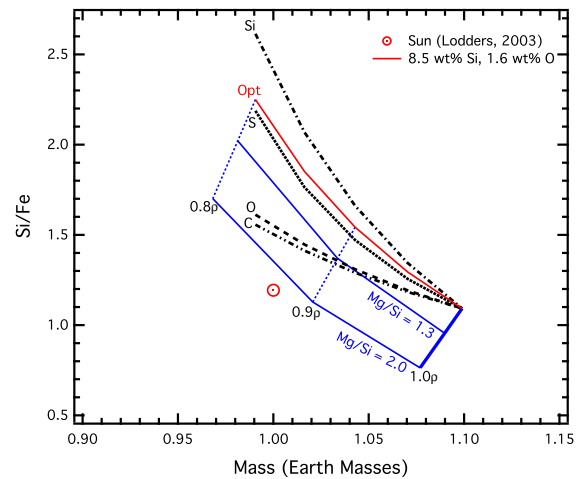}
\caption{Effect on composition and mass due to variation in mantle mineralogy, core size and core density deficit for a two-layer Earth-size planet (Fe-core + mantle). Changes in mantle Mg/Si are show in blue.  Variation in Si/Fe due to changes in the light element composition of the core is shown as dashed black lines, assuming only one element being responsible for the density deficit. Our proposed core light element composition of the molar ratio of 8.5 wt\% Si and 1.6 wt\% O is shown in red (Si/O$\sim$3). Si/Fe values for the Sun \citep{Lodd03} is included for reference.}
\label{NoUM}
\end{centering}
\end{figure}

\begin{figure}
\begin{centering}
\includegraphics[width=8cm]{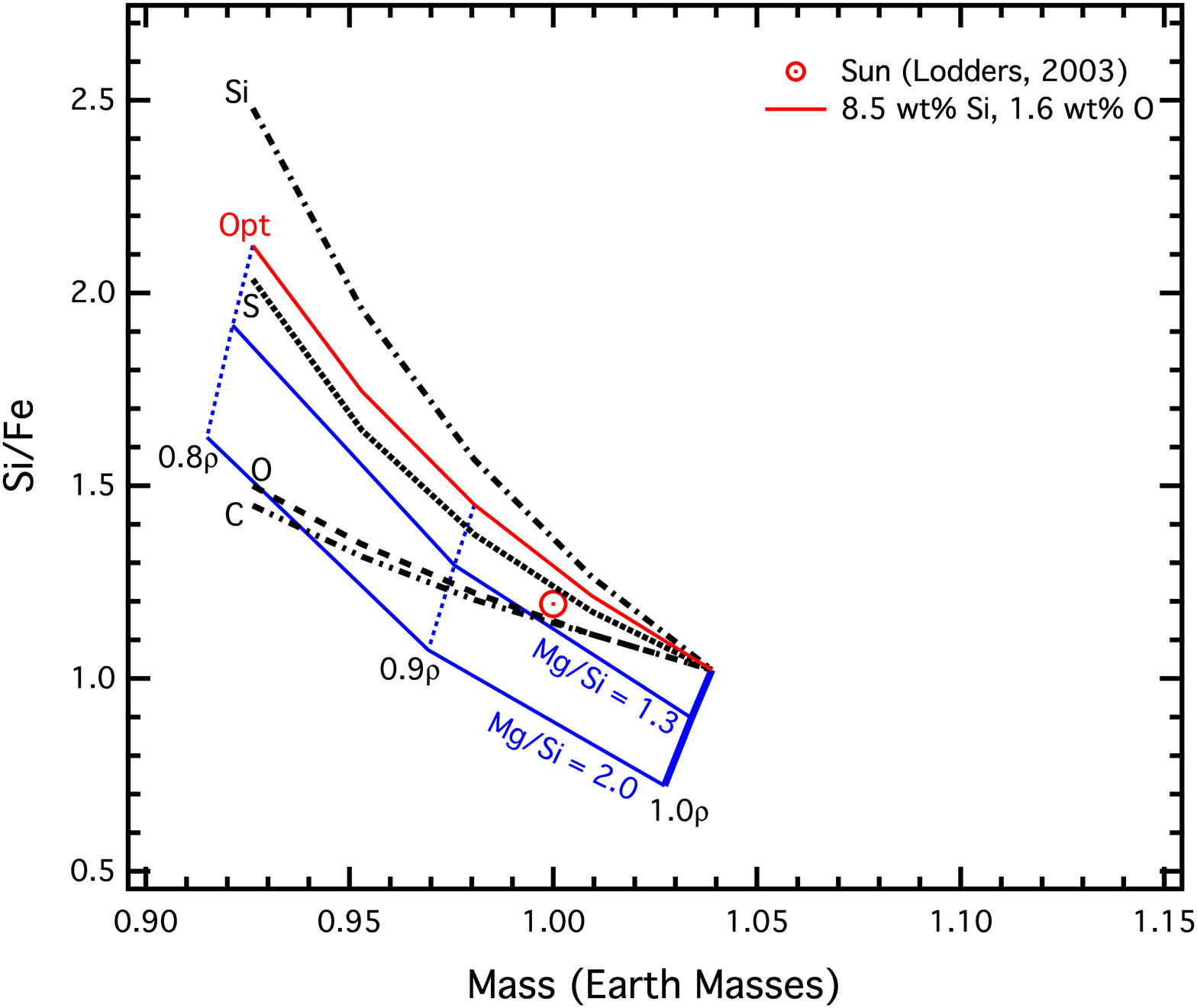}
\caption{Effect on composition and mass due to variation in mantle mineralogy, core size and core density deficit for a three-layer Earth-size planet (Fe-core + mantle + upper mantle). Changes in mantle Mg/Si are show in blue.  Variation in Si/Fe due to changes in the light element composition of the core is shown as dashed black lines, assuming only one element being responsible for the density deficit. Our proposed core light element composition of 8.5 wt\% Si and 1.6 wt\% O is shown in red. Si/Fe values for the Sun \citep{Lodd03} is included for reference.This figure is on the same scale as Figure \ref{NoUM} for comparison.}
\label{UM}
\end{centering}
\end{figure}

Of the parameters modeled here, we find core size, the density reduction due to the inclusion of an upper mantle and the presence of light elements in the core have the largest effects on the final mass of an Earth radius planet. Errors in Fe's bulk modulus, mantle composition, and mineralogy are secondary. There is degeneracy still present in our solutions however, with multiple model parameters and combination of model parameters producing planets of similar mass. As noted in \citet{Dorn15}, stellar host abundances of the main rock forming elements Mg, Si and Fe provide compositional constraints if assumed as proxies for planetary composition. This assumption stems from Solar abundances for these elements being within 5-10\% to those of a chondritic Earth \citep{Lodd10,McD03}. From stoichiometry and assuming these planets have a similar oxidation state to the Sun/Earth \citep{Unter14} as well as a fully differentiated planet, the ratios Si/Fe and Mg/Si provide sufficient constraints on the mass ratio of mantle to core and relative proportions of brigmanite and periclase.\\
\indent As the ratio of mantle to core dominates our determination of mass within mass-radius models, we adopt Si/Fe as a primary compositional constraint on planetary structure. We determine this ratio stoichiometrically by dividing the mass of the mantle or core by the molar mass of the constituent minerals/metals. As the density of the core decreases relative to pure-Fe, its molar mass decreases as well. The exact mass therefore depends on the molar mass of the light element partitioning into the core, with lighter elements (e.g. C, O) requiring incorporation of fewer moles and relatively higher mass elements (Si, S) requiring more in order to account for a given density deficit. We adopt the experimentally constrained core composition containing 8.5 wt\% Si and 1.6 wt\% O as a baseline core composition \citep[Figure \ref{NoUM}, red line, ][]{Fisch15}. Assuming no volume of solution, this composition results in a density deficit of $\sim$11\%.\\
\indent For a 1 R$_{\oplus}$ planet, we find that core density shifts the calculated mass with relatively small changes in Si/Fe. In contrast, the molar ratio of mantle minerals oppositely has only a small effect on total mass accompanying large shifts in Si/Fe (Figure \ref{NoUM}). In this parameter space only three parameters produce a planetary model which reproduces both the Earth's mass and Si/Fe: core radius, core density and Mg/Si of the mantle. Assuming a 1 CRE, two-layer planet model, no individual solution or combination of solutions reproduces both the Earth's mass and the Solar Si/Fe (Figure \ref{NoUM}). Including an upper mantle in our model, a planet of a core density deficit of $\sim$6\% and mantle Mg/Si $\sim$1.25 reproduces the Earth's composition and structure (Figure \ref{UM}). Such a model predicts a bulk composition of Mg/Fe and Si/Fe consistent with geochemical constraints, but depleted in Mg and Si relative to iron. We likewise reproduce the mass and radius of Venus if a slightly higher mantle Mg/Si of 1.35 is adopted (Table \ref{Kep36}). This is consistent with the slightly higher condensation temperature of olivine (Mg/Si = 2) versus pyroxene (Mg/Si = 1) \citep{Gros72}.\\
\indent In this simple, ``spherical cow,'' approach to modeling of an Earth sized planet interior, we find that there are three major parameters required to account for measured density and composition: the presence of light elements in the core, the radius of the core, and the inclusion of an upper mantle. Each of the other parameters modeled here contributes less than half with regards to density or composition.
\subsection{Application to Exoplanets}
Of our sensitivity study on the factors required to reproduce the Earth, the dominant factors of core fraction, light elements in the core, and consideration of upper mantle mineral phase transitions are independent of the details of terrestrial planet formation and chemistry. Therefore, we recommend that when modeling terrestrial exoplanets, models incorporate a 6\% density reduction of core materials due to incorporation of 8.5 wt\% Si and 1.6 wt\% O, and include the contribution to radius from the more low density upper mantle minerals. While this model inherently assumes that the mineralogies and core-compositions are inherently similar to Earth's, within this ``spherical cow'' model it allows us to compare apples-to-apples across different planetary systems and quantify how well these planets can be fit to a ``Earth-like'' model. Applying this method to other exoplanets is key to accurate description of the planet and determining their likelihood to be ``Earth-like.'' At 4.45$^{+0.33}_{-0.27}$ M$_{\oplus}$ and 1.486$\pm$0.035 R$_{\oplus}$ \citep{Cart12}, Kepler-36b is an ideal candidate for this analysis. Its mean density is 7.46$^{+0.74}_{-0.59}$ g/cm$^{3}$ and is therefore likely a terrestrial ``super Earth'' and its uncertainty in density (8\%) is among the smallest of this population of terrestrial planets. In the absence of stellar compositional constraints, we calculate a 3-layer model and determine the core radius that reproduces the mass of the planet with both a pure-Fe core or with a 6\% core density deficit due to light elements (Table \ref{Kep36}). Each aspect, in creating a more realistic planet, increases the core mass and radius of Kepler-36b, with a preferred model whose core has a radius and mass of 56.7\% and 39.8\% of the planet, respectively, compared to 54.7\% and 32.3\%, respectively, for the Earth. \\
\indent This proposed planetary structure has a Si/Fe ratio of 0.9, 10\% and 25\% smaller than the Earth and Sun respectively, consistent with Kepler-36's metallicity of [Fe/H] = -0.20$\pm$0.06 \citep{Adib12}. To date, though, the Si abundance of Kepler-36 has not been measured. Thus, in order to be consistent with our predicted composition based on mean density alone Kepler-36 must have a silicon abundance of [Si/H] = [Si/Fe]\footnote{[Si/Fe] = log(Si/Fe) - log(Si/Fe)$_{\odot}$} + [Fe/H] or $\sim$-0.32. More massive than the Earth, Kepler-36b is less likely to have suffered as significant silicate loss during early giant impacts \citep{Marc10}. Whether Kepler-36b then is indeed ``Earth-like,'' requires viewing our model in the broader context of what makes the Earth different from the other terrestrial planets in the Solar System.\\
\indent The chief difference between the Earth and the other terrestrial planets in our Solar system is that the Earth is a habitable, dynamic planet, with plate tectonics creating a deep water and carbon cycle. These cycles regulate surface C abundances \citep{Slee01}, and thus long-term climate, as well as providing a flux of water into the surface hydrosphere, all of which are vital to Earth being habitable. The driving force behind plate tectonics is the convection of the mantle. The fact that the Earth transports its interior heat via convection is due to a confluence of many factors, including composition, internal budget energy and planetary structure.\\
\indent While our model currently does not inform us regarding internal heat budget, it does provide constraints on both composition and structure, particularly the relative size of the convecting mantle, for an exoplanet of a given radius. To first order, the vigor of convection as determined by the system's Rayleigh number scales as the thickness of the convecting layer to the third power, with larger relative mantles convecting more vigorously than smaller ones of the same composition and interior heat budget \citep{Schu79, Schu80}. Planetary systems that are likely to have greater terrestrial heat budgets have been observed such that even with  with a relatively smaller mantle, these planets may still convect \citep{Unter15}. In the absence of spectroscopic measurements of heat producing elements Th, U, and $^{40}$K, however, this model provides first-order comparisons on the likelihood of mantle convection across planetary systems. \\
\begin{figure}
\begin{centering}
\includegraphics[width=8cm]{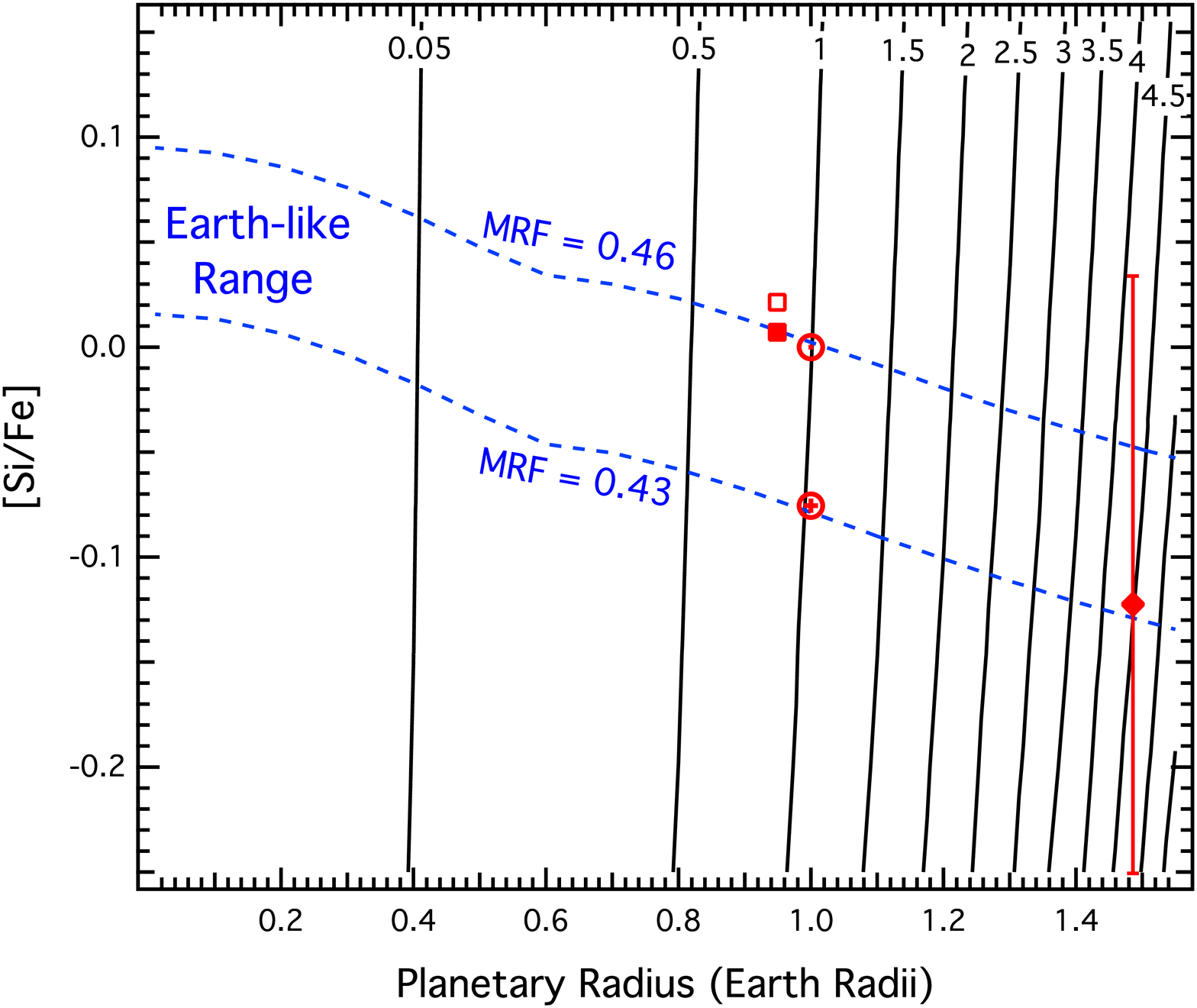}
\caption{Contours of constant mass (black, in Earth masses) and constant mantle radius fraction (MRF = (R - Core Radius)/R, blue-dashed) versus planetary radius and [Si/Fe]. Contours were calculated adopting a three-layer planetary model with an Fe-core with 6\% density deficit due to 8.5 wt\% Si and 1.6 wt\% O, brigmanite/periclase lower-mantle and pyroxene/olivine upper-mantle with Mg/Si = 1.25. Our calculated model for Kepler-36b is shown as a red diamond. The Earth and Venus (adopting Mg/Si = 1.25; open square, and Mg/Si = 1.35; solid square) are shown as well. A 1 R$\oplus$ planet of Solar composition ([Si/Fe] = 0) is also shown for comparison. The Solar model of \citet{Lodd03} was used to calculate relative [Si/Fe] abundances.}
\label{Grid}
\end{centering}
\end{figure}
With this in mind, we will restrict our definition of ``Earth-like,'' to those planets which are dynamically similar to the Earth, and thus have similar mantle fractions. Figure \ref{Grid} shows contours of constant mass and mantle radius fraction (MRF =  (R - Core Radius)/R) with respect to two observables: planetary radius and stellar [Si/Fe]. Contours were calculated using our three-layer model adopting a three-layer planetary model with an Fe-core with 6\% density deficit due to light element incorporation adopting the molar ratio of 8.5 wt\% Si and 1.6 wt\% O (Si/O$\sim$3), brigmanite and periclase lower-mantle and pyroxene and forsterite upper-mantle (Mg/Si = 1.25). Post-perovskite is the stable form of MgSiO$_{3}$ above 125 GPa and thus likely the dominant silicate phase in those planets with masses greater than the Earth. The density difference between brigmanite and post-perovskite, however, is less than 1.2\% \citep{Iita04}, and therefore including this transition in bridgmanite makes a minor difference compared to any of the primary factors we identify here. For the Earth (Si/Fe = 1), our model predicts a MRF = 0.43, whereas a planet of Solar composition (Si/Fe = 1.19) our model predicts a MRF = 0.46. Each of these values are very similar the actual value for the Earth of MRF = 0.45 \citep{PREM}. We adopt then, these MRF values as bounds for defining an ``Earth-like'' planet given this ``spherical cow'' approach. From this then, assuming a normal distribution in uncertainty, our model predicts Kepler-36b is only $\sim$20\% likely to be ``Earth-like'' in its structure and bulk composition given only its mass and radius. 

\section{Conclusion}
\begin{deluxetable}{lcccc}
\tablecolumns{5}
\tabletypesize{\scriptsize}
\tablewidth{0pt}
\tablecaption{Calculated core radius and mass fractions for Kepler-36b (4.45 M$_{\oplus}$, 1.486 R$_{\oplus}$) assuming a pure-Fe core and a 0.94$\rho$ Fe core containing a molar ratio of 8.5 wt\% Si and 1.6 wt\% O (Si/O$\sim$3), brigmanite/periclase lower mantle and both with and without and pyroxene/olivine upper mantles (Mg/Si = 1.25). Error bars are calculated using an uncertainty of 8\% in density manifesting entirely as an error in mass. Venus as calculated using this model is included for reference adopting Mg/Si = 1.25 and 1.35.}
\tablehead{\colhead{}&\colhead{core radius}&\colhead{core mass}&\colhead{Mg/Fe}&\colhead{Si/Fe}}
\startdata
Sun&-&-&1.22$^{*}$&1.19$^{*}$\nl
Earth&54.7\%$^{\dagger}$&32.3\%$^{\ddagger}$&1.11$^{\ddagger}$&1.0$^{\ddagger}$\nl
Venus$^{\mathsection}$&53.5\%&31.5\%&1.45&1.25 \nl
Venus (Mg/Si = 1.35)$^{\mathsection}$&53.5\%&31.5\%&1.50&1.21 \nl
\cutinhead{Kepler-36b: 100\% $\rho$ Fe}
no upper mantle&52.6$^{+4.4}_{-5.5}$\%&34.5$^{+7.2}_{-8.5}$\%&1.20$^{-0.32}_{+0.60}$&0.96$^{-0.25}_{+0.48}$\nl
&&&&\nl
upper mantle&54.0$^{+4.2}_{-5.3}$\%&37.3$^{+7.0}_{-8.4}$\%&1.06$^{-0.27}_{+0.50}$&0.85$^{-0.21}_{+0.39}$\nl
\cutinhead{Kepler-36b: 94\% $\rho$ Fe}
no upper mantle&55.3$^{+4.9}_{-6.0}$\%&36.9$^{+8.1}_{-9.2}$\%&1.00$^{-0.18}_{+0.74}$&1.14$^{-0.40}_{+0.35}$\nl
&&&&\nl
upper mantle&56.7$^{+4.9}_{-5.5}$\%&39.8$^{+8.7}_{-5.6}$\%&1.01$^{-0.29}_{+0.47}$&0.90$^{-0.23}_{+0.39}$\nl
\enddata
\tablerefs{\scriptsize $^{*}$\citet{Lodd03} $^{\dagger}$\citet{PREM} $^{\ddagger}$\citet{McD03}\nl $^{\mathsection}$Calculated using our model assuming 0.94$\rho$ Fe core due to the incorporation of a molar ratio of 8.5 wt\% Si and 1.6 wt\% O (Si/O$\sim$3) with brigmanite/periclase lower mantle and pyroxene/forsterite upper mantle (Mg/Si = 1.25)}
\label{Kep36} 
\end{deluxetable}
\indent While many calculated mass models presented here are well within the observational uncertainty of most terrestrial exoplanets, given longer time-series measurements these error bars will improve allowing us to more finely infer their interior structure. However, as we seek to discover ``Earth-like'' planets given this improved data, we must understand which aspects of a planet's composition and structure best constrain its mass-radius relationship. We find that small changes in the core size and composition, as well as the inclusion of an upper-mantle in mass-radius models have significant effects on the computed average density of a terrestrial planet. While degeneracy in determining the composition and interior structure still exists, the positive solution space is significantly reduced when compositional constraints are added when adopting the stellar, refractory element abundances as proxies for planetary composition, particularly Si/Fe. As we expand our definition of ``Earth-like'' beyond one merely of composition towards one based more on chemical and dynamic state, we can begin to limit which of these composition/structure solutions produce planets dynamically unlike Earth, i.e. lacking mantle convection and the potential for plate tectonics \citep{Unter14,Unter15}. Further degeneracy can be reduced by eliminating those core compositions that exceed the amount of light elements soluble in Fe, a topic of great interest in the mineral physics community. It is only with this solid foundation of understanding Earth that we can truly begin to extrapolate to higher mass-radius planets to determine their likelihood for being ``Earth-like.''
\acknowledgements 
This work is supported by NSF CAREER grant EAR-60023026 to WRP. Special thanks to Ian Rose for creating the initial mass-radius code for the CIDER 2014 Summer Institute. The Python scripts for Figure \ref{Earth_mod} available in latest release of the BurnMan software as planet\_builder.py with code for creating other figures available upon request. 
\bibliographystyle{apj}
\bibliography{Cow}

\begin{thebibliography}{}
\expandafter\ifx\csname natexlab\endcsname\relax\def\natexlab#1{#1}\fi

\bibitem[{Adibekyan {et~al.}(2012)Adibekyan, Sousa, Santos, Delgado~Mena,
  Gonz\'{a}lez~Hern\'{a}ndez, Israelian, Mayor, \& Khachatryan}]{Adib12}
Adibekyan, Sousa, S.~G., Santos, N.~C., {et~al.} 2012, Astronomy \&
  Astrophysics, 545, A32+

\bibitem[{Anderson {et~al.}(2001)Anderson, Dubrovinsky, Saxena, \&
  LeBihan}]{Ande01}
Anderson, O.~L., Dubrovinsky, L., Saxena, S.~K., \& LeBihan, T. 2001, Geophys.
  Res. Lett., 28, 399

\bibitem[{Anderson \& Ahrens(1994)}]{Ander94}
Anderson, W.~W., \& Ahrens, T.~J. 1994, Journal of Geophysical Research, 99,
  4273+

\bibitem[{Batalha {et~al.}(2011)Batalha, Borucki, Bryson, Buchhave, Caldwell,
  Christensen-Dalsgaard, Ciardi, Dunham, Fressin, Gautier, Gilliland, Haas,
  Howell, Jenkins, Kjeldsen, Koch, Latham, Lissauer, Marcy, Rowe, Sasselov,
  Seager, Steffen, Torres, Basri, Brown, Charbonneau, Christiansen, Clarke,
  Cochran, Dupree, Fabrycky, Fischer, Ford, Fortney, Girouard, Holman, Johnson,
  Isaacson, Klaus, Machalek, Moorehead, Morehead, Ragozzine, Tenenbaum,
  Twicken, Quinn, VanCleve, Walkowicz, Welsh, Devore, \& Gould}]{Bata11}
Batalha, N.~M., Borucki, W.~J., Bryson, S.~T., {et~al.} 2011, The Astrophysical
  Journal, 729, 27+

\bibitem[{Birch(1952)}]{Birch52}
Birch, F. 1952, Journal of Geophysical Research, 57, 227

\bibitem[{Birch(1961)}]{Birch61}
---. 1961, J. Geophys. Res., 66, 2199

\bibitem[{Carter {et~al.}(2012)Carter, Agol, Chaplin, Basu, Bedding, Buchhave,
  Christensen-Dalsgaard, Deck, Elsworth, Fabrycky, Ford, Fortney, Hale,
  Handberg, Hekker, Holman, Huber, Karoff, Kawaler, Kjeldsen, Lissauer, Lopez,
  Lund, Lundkvist, Metcalfe, Miglio, Rogers, Stello, Borucki, Bryson,
  Christiansen, Cochran, Geary, Gilliland, Haas, Hall, Howard, Jenkins, Klaus,
  Koch, Latham, MacQueen, Sasselov, Steffen, Twicken, \& Winn}]{Cart12}
Carter, J.~A., Agol, E., Chaplin, W.~J., {et~al.} 2012, Science, 337, 556

\bibitem[{Cottaar {et~al.}(2014)Cottaar, Heister, Rose, \& Unterborn}]{Cott14}
Cottaar, S., Heister, T., Rose, I., \& Unterborn, C. 2014, Geochem. Geophys.
  Geosyst., 15, 1164

\bibitem[{Dewaele {et~al.}(2006)Dewaele, Loubeyre, Occelli, Mezouar,
  Dorogokupets, \& Torrent}]{Dewa00}
Dewaele, A., Loubeyre, P., Occelli, F., {et~al.} 2006, Physical Review Letters,
  97, doi:10.1103/physrevlett.97.215504

\bibitem[{Dorn {et~al.}(2015)Dorn, Khan, Heng, Connolly, Alibert, Benz, \&
  Tackley}]{Dorn15}
Dorn, C., Khan, A., Heng, K., {et~al.} 2015, Astronomy \& Astrophysics, 577,
  A83+

\bibitem[{Driscoll \& Barnes(2015)}]{Dris15}
Driscoll, P.~E., \& Barnes, R. 2015, Astrobiology, 15, 739

\bibitem[{Dziewonski \& Anderson(1981)}]{PREM}
Dziewonski, A.~M., \& Anderson, D.~L. 1981, Physics of the Earth and Planetary
  Interiors, 25, 297

\bibitem[{Fischer {et~al.}(2015)Fischer, Nakajima, Campbell, Frost, Harries,
  Langenhorst, Miyajima, Pollok, \& Rubie}]{Fisch15}
Fischer, R.~A., Nakajima, Y., Campbell, A.~J., {et~al.} 2015, Geochimica et
  Cosmochimica Acta, 167, 177

\bibitem[{Foley(2015)}]{Fole15}
Foley, B.~J. 2015, The Astrophysical Journal, 812, 36+

\bibitem[{Fortney {et~al.}(2007)Fortney, Marley, \& Barnes}]{Fort07}
Fortney, J.~J., Marley, M.~S., \& Barnes, J.~W. 2007, The Astrophysical
  Journal, 659, 1661

\bibitem[{Grasset {et~al.}(2009)Grasset, Schneider, \& Sotin}]{Gras09}
Grasset, O., Schneider, J., \& Sotin, C. 2009, The Astrophysical Journal, 693,
  722

\bibitem[{Grossman(1972)}]{Gros72}
Grossman, L. 1972, Geochimica et Cosmochimica Acta, 36, 597

\bibitem[{Han {et~al.}(2014)Han, Wang, Wright, Feng, Zhao, Fakhouri, Brown, \&
  Hancock}]{ExoOrg}
Han, E., Wang, S.~X., Wright, J.~T., {et~al.} 2014, Publications of the
  Astronomical Society of the Pacific, 126, 827

\bibitem[{{Hatzes} {et~al.}(2011){Hatzes}, {Fridlund}, {Nachmani}, {Mazeh},
  {Valencia}, {H{\'{e}}brard}, {Carone}, {P{\"{a}}tzold}, {Udry}, {Bouchy},
  {Deleuil}, {Moutou}, {Barge}, {Bord{\'{e}}}, {Deeg}, {Tingley}, {Dvorak},
  {Gandolfi}, {Ferraz-Mello}, {Wuchterl}, {Guenther}, {Guillot}, {Rauer},
  {Erikson}, {Cabrera}, {Csizmadia}, {L{\'{e}}ger}, {Lammer}, {Weingrill},
  {Queloz}, {Alonso}, {Rouan}, \& {Schneider}}]{Hatz11}
{Hatzes}, A.~P., {Fridlund}, M., {Nachmani}, G., {et~al.} 2011, Astrophys. J.,
  743, 75+

\bibitem[{Iitaka {et~al.}(2004)Iitaka, Hirose, Kawamura, \& Murakami}]{Iita04}
Iitaka, T., Hirose, K., Kawamura, K., \& Murakami, M. 2004, Nature, 430, 442

\bibitem[{Irifune \& Isshiki(1998)}]{Irif98}
Irifune, T., \& Isshiki, M. 1998, Nature, 392, 702

\bibitem[{Javoy {et~al.}(2010)Javoy, Kaminski, Guyot, Andrault, Sanloup,
  Moreira, Labrosse, Jambon, Agrinier, Davaille, \& Jaupart}]{Javo10}
Javoy, M., Kaminski, E., Guyot, F., {et~al.} 2010, Earth and Planetary Science
  Letters, 293, 259

\bibitem[{Jeanloz(1979)}]{Jean79}
Jeanloz, R. 1979, J. Geophys. Res., 84, 6059

\bibitem[{Jeanloz \& Thompson(1983)}]{Jean83}
Jeanloz, R., \& Thompson, A.~B. 1983, Reviews of Geophysics, 21, 51

\bibitem[{Lodders(2003)}]{Lodd03}
Lodders, K. 2003, The Astrophysical Journal, 1220+

\bibitem[{Lodders(2010)}]{Lodd10}
---. 2010, in Principles and Perspectives in Cosmochemistry, ed. A.~Goswami \&
  B.~E. Reddy, Astrophysics and Space Science Proceedings (Springer Berlin
  Heidelberg), 379--417

\bibitem[{Lopez \& Fortney(2014)}]{Lope14}
Lopez, E.~D., \& Fortney, J.~J. 2014, The Astrophysical Journal, 792, 1+

\bibitem[{Marcus {et~al.}(2010)Marcus, Sasselov, Hernquist, \&
  Stewart}]{Marc10}
Marcus, R.~A., Sasselov, D., Hernquist, L., \& Stewart, S.~T. 2010, The
  Astrophysical Journal Letters, 712, L73

\bibitem[{McDonough(2003)}]{McD03}
McDonough, W.~F. 2003, {Compositional Model for the Earth's Core} (Elsevier),
  547--568

\bibitem[{Poirier(1994)}]{Poir94}
Poirier, J.-P. 1994, Physics of the Earth and Planetary Interiors, 85, 319

\bibitem[{Queloz {et~al.}(2009)Queloz, Bouchy, Moutou, Hatzes, H\'{e}brard,
  Alonso, Auvergne, Baglin, Barbieri, Barge, Benz, Bord\'{e}, Deeg, Deleuil,
  Dvorak, Erikson, Ferraz~Mello, Fridlund, Gandolfi, Gillon, Guenther, Guillot,
  Jorda, Hartmann, Lammer, L\'{e}ger, Llebaria, Lovis, Magain, Mayor, Mazeh,
  Ollivier, P\"{a}tzold, Pepe, Rauer, Rouan, Schneider, Segransan, Udry, \&
  Wuchterl}]{Quel09}
Queloz, D., Bouchy, F., Moutou, C., {et~al.} 2009, Astronomy and Astrophysics,
  506, 303

\bibitem[{Rudnick \& Gao(2003)}]{Rudn03}
Rudnick, R.~L., \& Gao, S. 2003, in Treatise on Geochemistry, ed. H.~D.
  Turekian (Oxford: Pergamon), 1--64

\bibitem[{Schubert(1979)}]{Schu79}
Schubert, G. 1979, Annual Review of Earth and Planetary Sciences, 7, 289

\bibitem[{Schubert {et~al.}(1980)Schubert, Stevenson, \& Cassen}]{Schu80}
Schubert, G., Stevenson, D., \& Cassen, P. 1980, Journal of Geophysical
  Research, 85, 2531+

\bibitem[{Seager {et~al.}(2007)Seager, Kuchner, Hier‐Majumder, \&
  Militzer}]{Seag07}
Seager, S., Kuchner, M., Hier‐Majumder, C.~A., \& Militzer, B. 2007, The
  Astrophysical Journal, 669, 1279

\bibitem[{Sleep \& Zahnle(2001)}]{Slee01}
Sleep, N.~H., \& Zahnle, K. 2001, Journal of Geophysical Research, 106, 1373+

\bibitem[{Sotin {et~al.}(2007)Sotin, Grasset, \& Mocquet}]{Sot07}
Sotin, C., Grasset, O., \& Mocquet, A. 2007, Icarus, 191, 337

\bibitem[{Stixrude \& Lithgow-Bertelloni(2005)}]{Stix05}
Stixrude, L., \& Lithgow-Bertelloni, C. 2005, Geophysical Journal
  International, 162, 610

\bibitem[{Swift {et~al.}(2012)Swift, Eggert, Hicks, Hamel, Caspersen,
  Schwegler, Collins, Nettelmann, \& Ackland}]{Swif12}
Swift, D.~C., Eggert, J.~H., Hicks, D.~G., {et~al.} 2012, The Astrophysical
  Journal, 744, 59+

\bibitem[{Unterborn {et~al.}(2015)Unterborn, Johnson, \& Panero}]{Unter15}
Unterborn, C.~T., Johnson, J.~A., \& Panero, W.~R. 2015, The Astrophysical
  Journal, 806, 139+

\bibitem[{Unterborn {et~al.}(2014)Unterborn, Kabbes, Pigott, Reaman, \&
  Panero}]{Unter14}
Unterborn, C.~T., Kabbes, J.~E., Pigott, J.~S., Reaman, D.~M., \& Panero, W.~R.
  2014, The Astrophysical Journal, 793, 124+

\bibitem[{Valencia {et~al.}(2006)Valencia, O'Connell, \& Sasselov}]{Vale06}
Valencia, D., O'Connell, R.~J., \& Sasselov, D. 2006, Icarus, 181, 545

\bibitem[{Valencia {et~al.}(2007{\natexlab{a}})Valencia, Sasselov, \&
  O'Connell}]{Vale07a}
Valencia, D., Sasselov, D.~D., \& O'Connell, R.~J. 2007{\natexlab{a}}, The
  Astrophysical Journal, 1413+

\bibitem[{Valencia {et~al.}(2007{\natexlab{b}})Valencia, Sasselov, \&
  O'Connell}]{Vale07b}
---. 2007{\natexlab{b}}, The Astrophysical Journal, 545+

\bibitem[{Vo\v{c}adlo(2007)}]{Voca07}
Vo\v{c}adlo, L. 2007, {Mineralogy of the Earth – The Earth's Core: Iron and
  Iron Alloys} (Elsevier), 91--120

\bibitem[{Wagner {et~al.}(2011)Wagner, Sohl, Hussmann, Grott, \&
  Rauer}]{Wagn11}
Wagner, F.~W., Sohl, F., Hussmann, H., Grott, M., \& Rauer, H. 2011, Icarus,
  214, 366

\bibitem[{Weiss \& Marcy(2014)}]{Weis14}
Weiss, L.~M., \& Marcy, G.~W. 2014, The Astrophysical Journal, 783, L6+

\bibitem[{Zapolsky \& Salpeter(1969)}]{Zapo69}
Zapolsky, H.~S., \& Salpeter, E.~E. 1969, The Astrophysical Journal, 158, 809+

\bibitem[{Zeng \& Sasselov(2013)}]{Zeng13}
Zeng, L., \& Sasselov, D. 2013, Publications of the Astronomical Society of the
  Pacific, 125, 227

\end{thebibliography}
\end{document}